\documentclass[twocolumn,english,pra,amsmath,amssymb,showpacs]{revtex4}
\usepackage[T1]{fontenc}
\usepackage[latin9]{inputenc}
\usepackage{amsmath}
\usepackage{graphicx}
\usepackage{amssymb}
\usepackage{esint}

\makeatletter
\@ifundefined{textcolor}{}
{%
 \definecolor{BLACK}{gray}{0}
 \definecolor{WHITE}{gray}{1}
 \definecolor{RED}{rgb}{1,0,0}
 \definecolor{GREEN}{rgb}{0,1,0}
 \definecolor{BLUE}{rgb}{0,0,1}
 \definecolor{CYAN}{cmyk}{1,0,0,0}
 \definecolor{MAGENTA}{cmyk}{0,1,0,0}
 \definecolor{YELLOW}{cmyk}{0,0,1,0}
 }

\makeatother

\usepackage{babel}

\begin{document}

\title{Tunneling Qubit Operation on a Protected Josephson Junction Array}

\author{Zhi Yin}

\affiliation{Zhejiang Insitute of Modern Physics, Zhejiang University, Hangzhou
310027, China}

\author{Sheng-Wen Li}

\affiliation{Zhejiang Insitute of Modern Physics, Zhejiang University, Hangzhou
310027, China}

\author{Yi-Xin Chen}

\email{yxchen@zimp.zju.edu.cn}

\affiliation{Zhejiang Insitute of Modern Physics, Zhejiang University, Hangzhou
310027, China}

\begin{abstract}
We propose a complete quantum computation process on a topologically
protected Josephson junction array system, originally proposed by
Ioffe and Feigel\textquoteright{}man {[}Phys. Rev. B 66, 224503 (2002){]}.
Logical qubits for computation are encoded in the punctured array.
The number of qubits is determined by the number of holes. The topological
degeneracy is lightly shifted by tuning the flux along specific paths.
We show how to perform both single-qubit and basic quantum-gate operations
in this system, especially the controlled-NOT (CNOT) gate.
\end{abstract}
\maketitle

\section{Introduction}

Quantum algorithm shows us a tempting future to conquer some problems
that are hard to solve using classical computers\cite{nielson}, such
as, factorizing large numbers\cite{shor99}, searching in a large
database\cite{gro97}, and so on. However, the physical realization
of a quantum computer turns out to be a really difficult work. The
greatest obstacle is the decoherence of system induced by unavoidable
coupling with the surrounding environment\cite{zur03}. Since quantum
devices work in a microscopic scale, their interactions with the environment
become fatal to the quantum coherence of the system. Someting has
to be done to increase the decoherence time so as to complete enough
operations within so short a peroid of time.

Therefore, error correcting code (ECC) is needed\cite{nielson}. With
the help of ECC a small error rate can be tolerated. Howerver, since
the necessary error-rate threshold still lies far beyond the present
available technology, other methods to fight against error should
be considered. One promising method is topological quantum computation
(TQC) \cite{kit03}\cite{pres}\cite{nay08}.

Topological order\cite{wen}\cite{wen90}\cite{wen95}\cite{wen02}
is a new kind of quantum order. Although the mechanism of topological
order is not quite clear, we can find the following features in a
system in topological order:
\begin{enumerate}
\item The degeneracy of the ground state closely relates to the topology
of the physical system (usually 2-dimensional).
\item The system is immune to local perturbation and this property can be
used to protect qubits.
\item Excited quasiparticles may obey anyon statistics (Abelian or non-Abelian)\cite{wen}.
\end{enumerate}
With its great advantages, non-Abelian TQC\cite{nay08} attracts us
the most. In a non-Abelian topological phase, not only can the encoded
qubits be protected, but also the operation process can be protected
intrinsically through the braiding of non-Abelian anyons. Much has
been discussed about the Fibonacci anyons that can be used for TQC
directly in Quantum Hall Effect. However it is hard to control anyons
in real quantum Hall systems. Besides, Kitaev proposed a spin-1/2
honeycomb model\cite{kit06}, which seems achievable, followed by
some different possible physical implementations\cite{duan03}\cite{jia08}\cite{zha07}\cite{you08}\cite{vi08}.
The work in Ref.\cite{vi08} proposed a more sophisticated method
for the anyon detection process. Unforturnately, the non-Abelian anyons
that emerged in this model do not act well enough to complete the
full braiding operation of TQC. Attractively, B. Doucot, \textit{et
al.} gave a complete description of the rhombus Josephson junction
array (JJA) with Abelian and non-Abelian gauge theories \cite{doucot04}.
However, this is too complex for implementation.

Because of all these difficulties we are facing, we will turn to the
Abelian topological phase in which qubits can be protected as well.
Recently, Kou showed how to do computations in the toric code model
by using the tunneling effect theoretically \cite{kou09prl}\cite{kou09ar4}\cite{kou09ar5}.
However, the original toric code model calls for four-body interaction,
which is hard to implement in real systems.

The Josephson junction seems to one of the candidates to implement
a realistic quantum computation \cite{doucot02}\cite{bla01}\cite{sergey08}\cite{iof02}.
Ioffe and Feigel'man proposed a hexagonal Josephson junction lattice
that showed some topological protecting properties which can be used
as quantum memory \cite{iof02}.

Based on their work, we propose a complete TQC scheme in this article.
We focus on encoding logical qubits in the array and the construction
of quantum gates. Similar to the method of Kou, we also show how to
implement fundamental quantum gates, especially the \textsc{CNOT}
gate.

The article is organized as follows. In Sec. II, we introduce Ioffe
and Feigel'man's model in detail. In Sec. III, we study the ground-state
behavior. Topological degeneracy is used to encode qubits in the array.
In Sec. IV we show how to manipulate the qubits to complete basic
quantum gate operation. Finally, we draw summary in Sec. V.

\section{Hexagonal Josephson Lattice}

To describe the computation process we first introduce the model proposed
by Ioffe and Feigel'man \cite{iof02} in this section. In this work,
They showed that the array system exhibits some topological protecting
properties and thus can be used as quantum memory. We discuss how
to do computations based on their work in the next section. Now we
recount some basic properties of their model in detail first. These
properties are essential for the construction of computation gates.
For the convenience of the computation we talk about, we make a small
modification with the boundary and discuss further the quasiparticle
excitation, especially the vortex, which agrees, in essence, with
the original model.

First, we give the basic building element and explain how it works.
Then we show how the array is arranged and give the dynamics of the
system. Finally, we talk about the topological degeneracy of the ground
space and the low-lying excitations.

\subsection{Basic rhombus}

\begin{figure}
(a) %
\begin{minipage}[c]{1in}%
\includegraphics[width=1in]{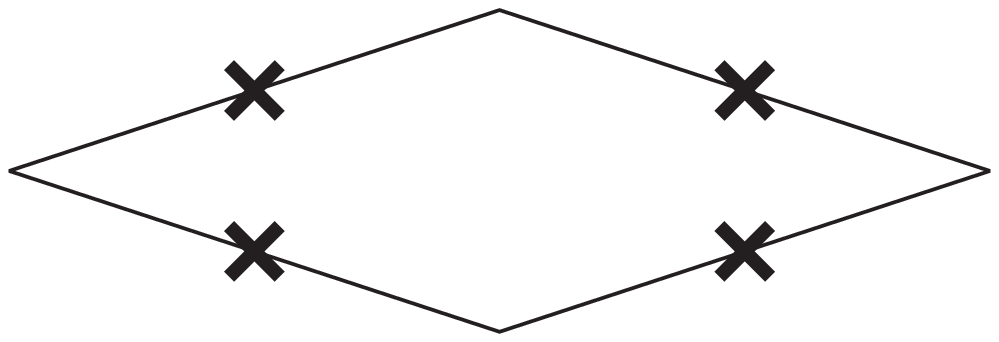}%
\end{minipage}\hspace{0.5in}(b) %
\begin{minipage}[c]{1.2in}%
 \includegraphics[width=1.2in,height=0.9in]{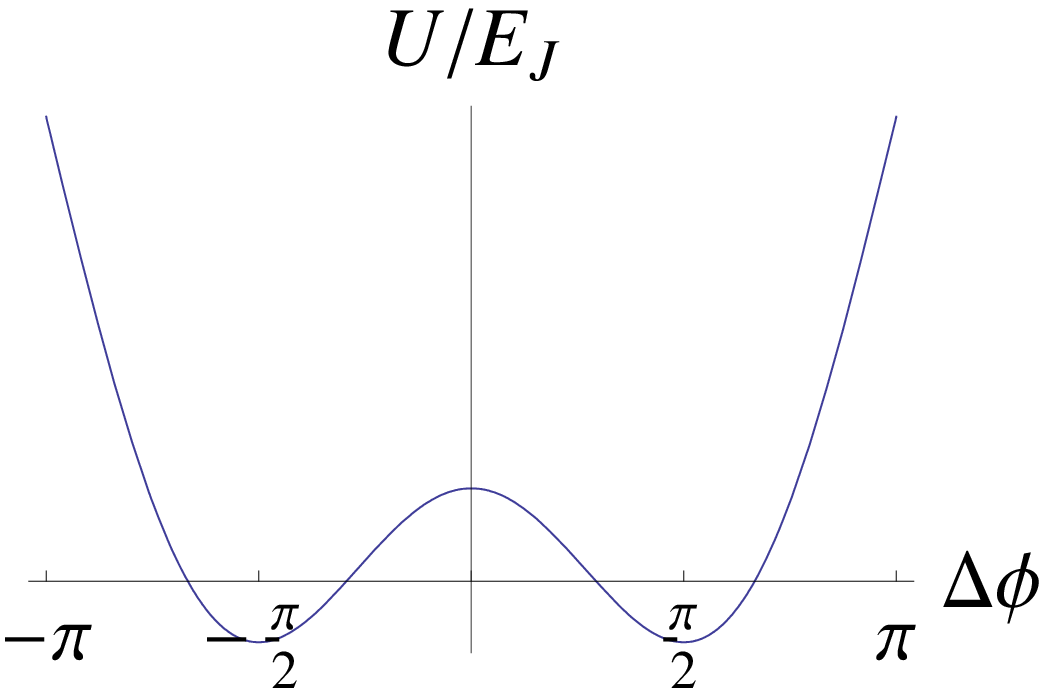}%
\end{minipage}\caption{(Color online) Construction of the basic element in the array. Four
identical Josephson junctions form a closed loop, embedding flux $\Phi_{0}/2$.
The classical potential is shown on the right. }

\label{potential}
\end{figure}

As shown in Fig.\ref{potential}, the basic building block is a rhombus
made of a four-junction superconducting loop placed in a magnetic
field. The flux through the rhombus is $\Phi_{0}/2$, where $\Phi_{0}=h/2e$
is the flux quanta. We require that all the Josephson junctions are
the same and work in the {}``phase regime'', that is, $E_{J}\gg E_{C}$,
where $E_{J}=(\hbar/2e)I_{c}$ and $E_{C}=e^{2}/2C$.

This system, as is much discussed in the literature, forms a simple
two-level {}``flux qubit'' (or persistent-current qubit)\cite{bla01}\cite{orl99}.
In the following, we briefly explain how the qubit system emerges.

We denote the phase of the superconductor islands with $\psi_{i}$.
The Josephson system is described with the Lagrangian \begin{equation}
\mathcal{L}=\sum_{\langle ij\rangle}\frac{\hbar^{2}}{16E_{C}}(\dot{\psi}_{i}-\dot{\psi}_{j})^{2}+E_{J}\cos(\psi_{i}-\psi_{j}-a_{ij}),\label{lagrangian}\end{equation}
 where $a_{ij}=\frac{2\pi}{\Phi_{0}}\int_{i}^{j}\vec{A}\cdot d\vec{l}$.
The quantity in the cosine term is the the gauge-invariant phase difference
across each junction.

Here, we make an approximation that three of the junctions have the
same phase difference. As all phase differences should sum up to $2\pi\Phi_{ext}/\Phi_{0}$,
we get \begin{equation}
U=4E_{J}-3E_{J}\cos\frac{\Delta\phi}{2}+E_{J}\cos\frac{3\Delta\phi}{2},\label{rhoU}\end{equation}
 where $\Delta\phi$ indicates the phase difference between the two
far ends of the rhombus. Neglecting the kinetic part, this energy
has two equivalent classical minima at $\Delta\phi=\pm\pi/2$ (as
shown in Fig.\ref{potential}), that is, the phase difference across
each junction is $\pm\pi/4$. So the two classical ground states correspond
to states in which there are both clockwise and counterclockwise supercurrents
in the loop. This is the reason why we call it a persistent-current
qubit. We denote the two states as $|\uparrow\rangle$ and $|\downarrow\rangle$.
Quantum fluctuation makes the electron tunnel between the two classical
states in the wells and that splits the degenerate ground states.
The low-energy configuration can be regarded as a two-level qubit
system. The effective Hamiltonian is \begin{equation}
H=\tilde{t}\sigma_{x},\label{rhoH}\end{equation}
 where $\tilde{t}\sim\sqrt{E_{J}E_{C}}e^{-S_{0}}$, and $S_{0}$ is
the classical action taking the path between the two minima. The value
of tunneling energy is estimated regarding the potential as quadratic.

Let us see what happens when the condition is slightly changed from
the ideal. Notice that the dynamics require that the flux through
the rhombus is \emph{exactly} half a flux quanta. If the external
flux is a bit different from the ideal value, $\Phi_{0}/2+\delta\Phi$,
we add a first order change to the potential, \begin{equation}
\begin{split}\tilde{U} & =4E_{J}-3E_{J}\cos\frac{\Delta\phi}{2}-E_{J}\cos(\pi+2\pi\frac{\delta\Phi}{\Phi_{0}}-\frac{3\Delta\phi}{2})\\
 & \simeq U+E_{J}\sin\frac{3\Delta\phi}{2}\cdot2\pi\frac{\delta\Phi}{\Phi_{0}}.\end{split}
\end{equation}
As a result of the slight change of the flux, the two minima of the
double well are no longer degenerate but separated by a gap $\Delta E=2\sqrt{2}\pi E_{J}\delta\Phi/\Phi_{0}$.
Then the effective Hamiltonian is $\tilde{H}=\tilde{t}\sigma_{x}+\frac{1}{2}\Delta E\sigma_{z}$.

This is important for operations. We can control the evolution of
the system by adjusting the flux through the rhombus. Below, in the
Josephson array building from the basic rhombi, logical qubits are
encoded for computation. Some gate operations are done just by the
method of tuning the flux of the rhombi along specific paths.

\subsection{Array}

\begin{figure}
\begin{centering}
(a)\includegraphics[width=3in]{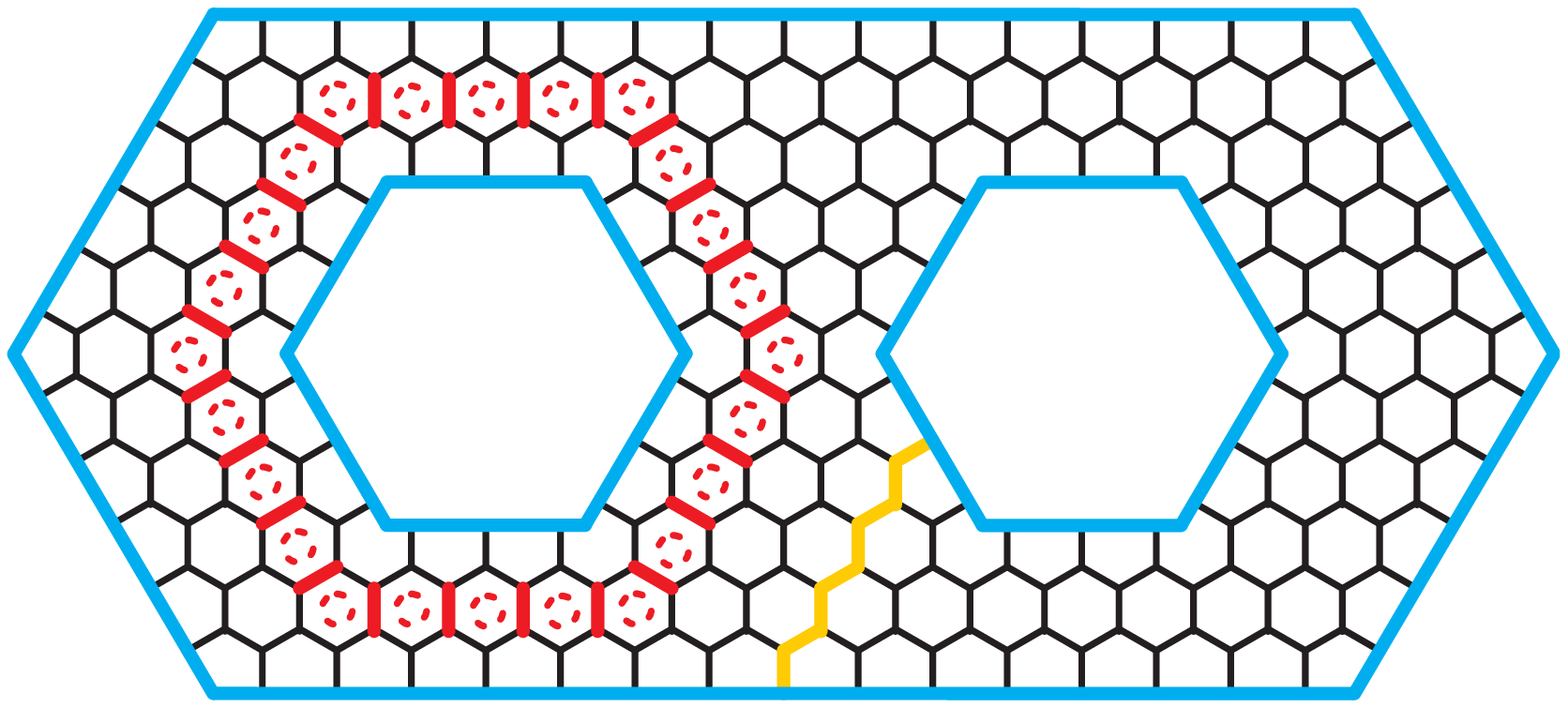} \vspace{0.2in}

\par\end{centering}

\begin{centering}
(b) \includegraphics[width=3in]{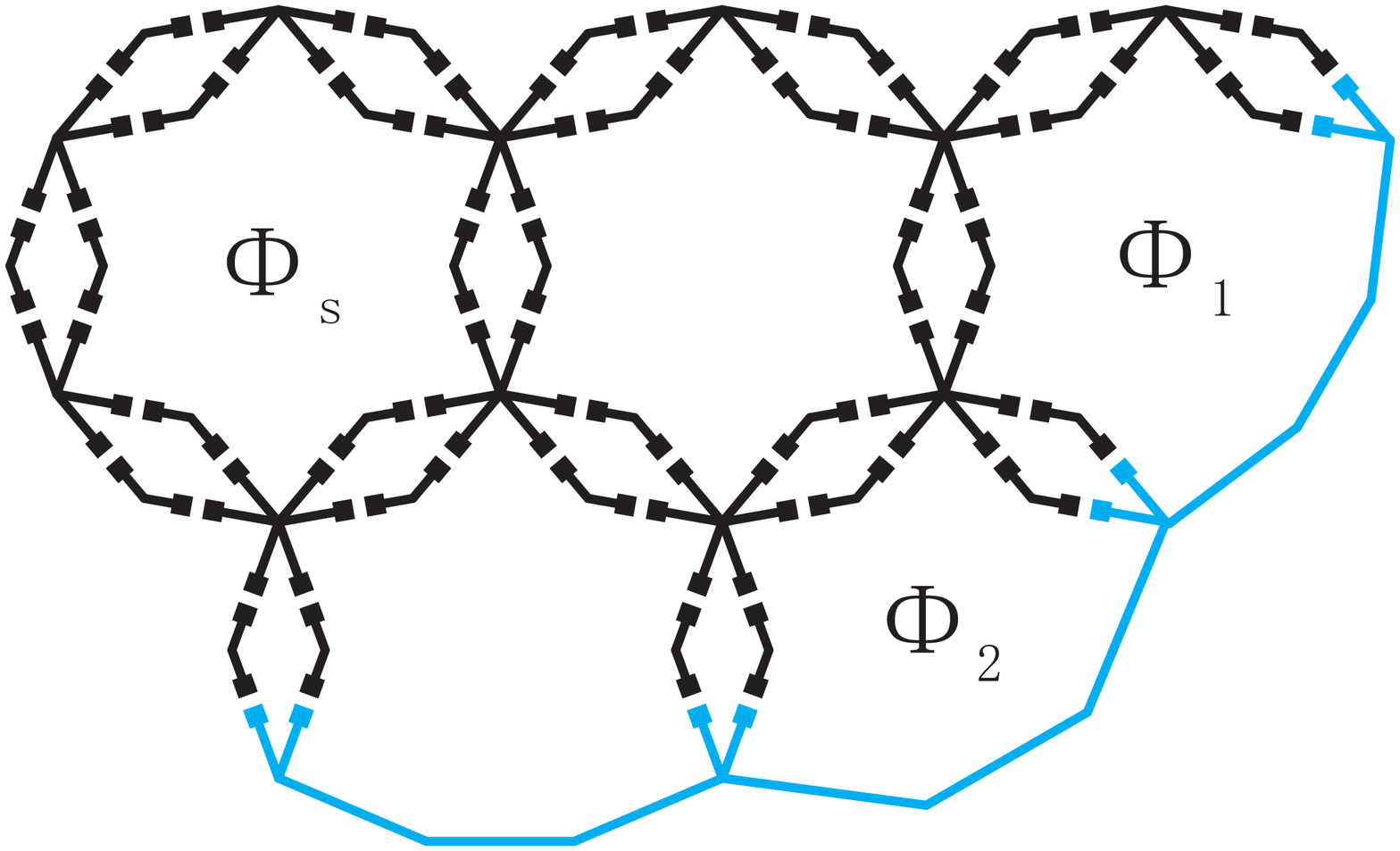}
\par\end{centering}

\caption{(Color online) Construction of the array with the basic rhombus. Six
rhombi form a hexagon. Each hexagram embeds a flux of $\Phi_{s}=(n_{s}+\frac{1}{2})\Phi_{0}$.
The array is punctured with several holes. The number of the holes
relates to the degeneracy of the ground space. The boundary, as shown
in the lower figure and indicated by a (blue) solid line, is a whole
superconductor wire connecting all the rhombi along the boundary.
The flux through the polygon loops on the boundary are $\Phi_{1}=n_{1}\Phi_{0}$
and $\Phi_{2}=(n_{2}+\frac{3}{4})\Phi_{0}$. }

\label{array}
\end{figure}

Now we have got the bricks we need. In this part we show how to build
the Josephson array system.

The basic rhombi are arranged as shown in Fig.\ref{array}. For simplicity,
a rhombus is indicated by a short line in the figure. Six rhombi form
a hexagon. The array is placed in a uniform magnetic field. We should
carefully tune the shape of the rhombi, so that the flux through the
hexagram in the center is a half-integer multiple of $\Phi_{0}$,
$\Phi_{s}=(n_{s}+1/2)\Phi_{0}$. The centers of the hexagons are denoted
with $a,b,c,$ and so on and individual rhombi between the corresponding
two centers with $(ab),(bc)$ .... The array contains $K$ holes.
Soon we will show that, in fact, each hole can be treated as a single
logical qubit.

In the array, the rhombi form loops, so we should check the gauge-invariant
condition over again. To satisfy the constraint condition for all
the closed loops in the array, the sum of the phase differences around
each hexagon should equal $2\pi\Phi_{s}/\Phi_{0}$ {[}i.e., $(2n_{s}+1)\pi${]}.
This is well consistent with the configuration that all the $\Delta\phi_{ab}$
of the six rhombi on the edges of a hexagon take the ground-state
value $\pi/2$ (or all take $-\pi/2$). The flip of a single rhombus
is not allowed because it changes the sum of the phase differences
around the loop by $\pi$ and violates the constraint. However, we
can flip rhombi of an even number in a hexagon at the same time (e.g.,
two, four or six). Choosing the two ground states as a basis, we can
write the constraint in the operator form, \begin{equation}
\hat{P}_{a}=\prod_{b}\sigma_{ab}^{z}=\mathbf{1}.\label{constraint}\end{equation}
 The product takes the rhombi on the hexagon loops.

Because of the existence of the constraint, the rhombi embedded in
the array can no longer behave as in Eq.(\ref{rhoH}), which does
not commute with the constraint. The simplest dynamics compatible
with constraint contain flips of three adjacent rhombi, $\hat{Q}_{abc}=\sigma_{ab}^{x}\sigma_{bc}^{x}\sigma_{ca}^{x}$,
such that two rhombi flip at the same time in the corresponding three
hexagons. This gives the main contribution to the Hamiltonian of the
system \begin{equation}
\hat{\mathcal{H}}_{0}=-r\sum_{abc}\hat{Q}_{abc}.\label{aryH}\end{equation}
 Here, $r\sim\sqrt{E_{J}E_{C}}\exp(-3S_{0})$ is estimated to be analogous
to $\tilde{t}$ in the last section. The action in the exponential
term is three times of that of a single rhombus because of the simultaneous
flip of the three rhombi.

\subsection{Boundary}

Now we talk about the boundary. We should treat it carefully. The
boundary here is arranged a bit differently from Ioffe's original
model. Different boundary conditions of the toric code have been discussed
in detail\cite{den02}. In the system considered here, similarly,
the boundary is arranged more like the {}``rough edge''.

Here, we arrange the boundary rhombi in such a way that they behave
like those in the bulk, as described by Eq.(\ref{aryH}). Each rhombus
on the boundary still stays close to the other two. A superconductor
line connects the unattached rhombi at the outer and inner boundary
(like the blue line in Fig.\ref{array}).

The geometry of the polygon loops on the boundary should be carefully
treated. A boundary loop has $m$ rhombi, usually three or four. We
want to maintain the dynamics and constraint of the form as before,
that is, two rhombi must flip at one time. So the flux embedded should
be consistent with the sum of the phase differences of the $m$ rhombi,
i.e., $(n_{b}+\frac{m}{4})\Phi_{0}$, where $n_{b}$ is an integer.
Then the Hamiltonian Eq.(\ref{aryH}) is still kept for the rhombi
on the boundary.

We complete the construction of the Josephson array and give the low-energy
dynamics. In the following section, we will discuss the properties
of the ground-state space and excitations in detail.

\subsection{Topological degeneracy, charge and vortex excitation}

In this section, we discuss the properties of the ground space and
excitations. The topological degeneracy of the ground space can be
used to encode logical qubits for computation. We show two kinds of
quasiparticles in the system, charge and vortex. We can operate qubits
by controlling their motion in the array.

First, let us find the ground state of the system. Ignoring the constraint,
we can easily get the ground state of Hamiltonian (\ref{aryH}) as
$|O\rangle=\prod|+\rangle=\prod(|\uparrow\rangle+|\downarrow\rangle)/\sqrt{2}$.
Of course, it is illegal and incompatible with the constraint. We
can project it into the physical space with the help of an operator
\begin{equation}
\hat{\mathcal{P}}=\prod_{a}\frac{1+\hat{P}_{a}}{\sqrt{2}}.\end{equation}
 $|g\rangle=\hat{\mathcal{P}}|O\rangle$ is a ground state of the
system, but not the only one.

To find all the ground states, we can turn to the help of a path operator
\begin{equation}
\hat{T}_{\gamma_{k}}=\prod^{(\gamma_{k})}\sigma_{ab}^{x},\end{equation}
 where $\gamma_{k}$ takes the big (red) loop in Fig.\ref{array}.
Notice that all the $\hat{T}_{\gamma_{k}'}$ with the paths homotopic
with $\gamma_{k}$ are equivalent. Also it is easy to check that $\hat{T}_{\gamma_{k}}$
commutes with both $H$ and $\hat{P}_{a}$. So obviously $\hat{T}_{\gamma_{k}}|g\rangle$
is another ground state of $H$. Noting that $\hat{T}_{\gamma_{k}}^{2}=\mathbf{\mathbf{1}}$,
we can construct all the ground states \begin{equation}
|G\rangle=\prod_{\gamma_{k}}\frac{1\pm\hat{T}_{\gamma_{k}}}{\sqrt{2}}\hat{\mathcal{P}}|O\rangle.\end{equation}
 Here, the product contains all the $K$ {}``basic loops'' (i.e.,
loops around each single hole).

Note that in the representation above, $\hat{T}_{\gamma_{k}}|G\rangle=\pm|G\rangle$.
So we can interpret $\hat{T}_{\gamma_{k}}$ as the operator measuring
the {}``spin'' direction of the $k$th hole. We will utilize this
physical meaning of $\hat{T}_{\gamma_{k}}$ for encoding in the next
section.

Now we find the whole ground space of the array system with $K$ holes.
The dimension of the ground space is just $2^{K}$. It is interesting
that the degeneracy correlates with the topology of the system. This
is why it is called topological degeneracy. We can encode $K$ qubit
in this space and do computations in it.

Based on the knowledge of the ground space, let us study the low-lying
excitations of the system. An open string product of $\sigma^{z}$,
which is compatible with the constraint, excites two {}``charges''
at the two ends of the string. With the help of the anti-commutation
relation with the Hamiltonian, after a simple algebra computation,
we can deduce that the energy of each single {}``charge'' is $2r$
above the ground.

Unfortunately, a string product of $\sigma^{x}$ is not allowed by
the constraint, which is different from the case in toric code model\cite{kit03}.
To get another type of excitation, we should break the constraint
Eq.(\ref{constraint}) and contemplate the problem from the original
Lagrangian of the array. Approximately, we only keep the degree of
freedom of the superconductor islands at the two ends of the rhombi
(i.e., at the vertices of the hexagons). Correspondingly, we use the
modified potential Eq.(\ref{rhoU}). Then the Lagrangian of the array
is \begin{equation}
\mathcal{L}=\sum_{\langle ij\rangle}\frac{\hbar^{2}}{16E_{C}}(\dot{\psi}_{i}-\dot{\psi}_{j})^{2}+3E_{J}\cos\frac{\Delta\psi_{ij}}{2}-E_{J}\cos\frac{3\Delta\psi_{ij}}{2},\end{equation}
 where $\Delta\psi_{ij}=\psi_{i}-\psi_{j}-\int_{i}^{j}\vec{A}\cdot\vec{l}$.
Taking the continuum limit, the Josephson array system can be mapped
to a Type-II superconductor\cite{orl91}. A vortex solution exists
in the system, taking one flux quanta. The energy of a single vortex
is \begin{equation}
E_{g}=\frac{\Phi_{0}^{2}}{4\pi\mu_{0}\lambda^{2}}\ln\frac{\lambda}{\xi}.\end{equation}
 Here, $\lambda^{2}=\Phi_{0}^{2}/9\pi^{2}\mu_{0}E_{J}$, and $\xi$
is the length between the two centers of the hexagons.

Vortices stay at the hexagons where the constraint is broken. The
dynamics there are changed correspondingly. There is one kind of dynamic
that interests us most, the single flip of a rhombus. To get this
kind of dynamics, we carefully change the flux of two adjacent hexagrams
to integer multiples of $\Phi_{0}$. Two vortices appear and stay
here. Single flip happens only at the rhombus between the two hexagons.
Since the other rhombi stay close to the hexagons that satisfy the
constraint, the single flip is still not allowed. This is also the
reason why we must change the flux of two adjacent hexagons, but not
one or two separated hexagons. If we change the flux of only one hexagon,
we still cannot get the single flip because it is forbidden by the
adjacent hexagon.

Now we can add a $\sigma^{x}$ term to $\hat{\mathcal{H}}_{0}$ with
the amplitude $\tilde{t}$ at the rhombus between the two adjacent
vortices. Or rather, we can interpret it in such a way that two vortices
with each energy $E_{g}$ are excited by a $\sigma^{x}$ operation.
This term is important for operations that we will discuss later.
So we have got two types of quasi-particle excitations in the array.

\section{Protected Qubit and Computation}

After all the discussions about the properties of the array we focus,
in this section, on the computation process. It was mentioned in Ref.
\cite{iof02} that motion of quasi-particles along topologically non-trivial
paths changes the state. We care about how to achieve it. In this
section, we show how to encode logical qubits in the array, and propose
an exact method to operate qubits in lab.

However, first we give a short review of stabilizer code formalism.
In fact, this system can be regarded as stabilizer code. Then we show
how the topological degenerated ground space is protected from local
noise. Logical qubits are encoded in the {}``holes''. The spin directions
of holes can be used to denote the state of the $K$ logical qubits.
Finally, we show how to implement the single qubit and multi-qubits
operations.

\subsection{Stabilizer code}

Stabilizer code is developed as a general theory of quantum error-correction
code\cite{nielson}. In this formalism, $N$ physical qubits are used
to encode $K$ logical qubits ($N>K$). Certain types of errors can
be detected and corrected so that the reliability of quantum computation
is greatly improved.

First, we define the Pauli group $G_{N}$ on $N$ qubits as \begin{equation}
G_{N}=\{i^{k}(\sigma_{1}^{x})^{a_{1}}(\sigma_{1}^{z})^{b_{1}},\ldots,(\sigma_{N}^{x})^{a_{N}}(\sigma_{N}^{z})^{b_{N}}\},\end{equation}
 where $k,a_{i},b_{i}$ are integers. $G_{N}$ contains all the $N$-body
Pauli operators with coefficients $\pm1,\pm i$. Note that any two
elements in the Pauli group either commute or anti-commute with each
other.

A subgroup $S$ of $G_{N}$ is called a \emph{stabilizer group} if
there exists a nontrivial space $V_{S}$ stabilized by $S$, \begin{equation}
V_{S}=\{|\psi\rangle|\quad g|\psi\rangle=|\psi\rangle,\quad\forall g\in S\}.\end{equation}
 It can be proved that $V_{S}$ is $2^{K}$-dimensional if the group
$S$ has $N-K$ independent generators. We denote the stabilizer group
as $S=\langle g_{1},g_{2},\ldots,g_{N-K}\rangle$, where $g_{i}$
is one of the $N-K$ generators of the group. The elements in $S$
are called \emph{stabilizer operators}, or just \emph{stabilizers}.

$K$ logical qubits are encoded in the stabilized space $V_{S}$.
To correct errors, we should first notice that the effects of errors
can be represented by the one-body elements in the Pauli group $G_{N}$,
because errors are usually local. When no error happens, after measuring
all the stabilizers in $S$ (just measuring all the $N-K$ generators
is ok.), all results are 1. Assume one error, denoted by $E\in G_{N}-S$,
happens, so that the original computation state $|\psi\rangle\in V_{S}$
is changed to $E|\psi\rangle$. If $E$ anti-commutes with at least
one stabilizer in $S$, e.g., $g$, the measurement of $g$ gives
the result $-1$, then we know the error type and we can correct it
by corresponding operation. However, there may exist some elements
in $G_{N}-S$ that also commute with all the elements in $G_{N}$.
This type of errors can not be detected or corrected.

The basis of logical qubits should be written out for computation.
We can always find $K$ operators $\bar{Z}_{i}$ from $G_{N}-S$ such
that $g_{1},\ldots,g_{N-K},\bar{Z}_{1},\ldots,\bar{Z}_{K}$ form an
independent operator set. $\bar{Z}_{i}$ plays the role of a logical
operator $\sigma_{z}$. The logical computational basis $|x_{1},x_{2}\ldots x_{K}\rangle_{L}$
is actually the state stabilized by the group \begin{equation}
\langle g_{1},\ldots,g_{N-K},(-1)^{x_{1}}\bar{Z}_{1},\ldots,(-1)^{x_{K}}\bar{Z}_{K}\rangle.\end{equation}
 The logical $\bar{X}_{i}$ can be also found from $G_{N}-S$ satisfying
:
\begin{enumerate}
\item $[\bar{X}_{i},\bar{X}_{j}]=0$ and
\item $\{\bar{X}_{i},\bar{Z}_{j}\}=2\delta_{ij}$.
\end{enumerate}
We emphasize here that the logical operator set $\{\bar{Z_{i}}\}$
is not unique. Also, we did not limit $\bar{Z}_{i}$ to be of some
certain form such as a product of several $\sigma^{z}$. It can be
a mixed product of $\sigma^{z}$ and $\sigma^{x}$, or even a product
that contains only $\sigma^{x}$. The only condition required is that
$g_{1},\ldots,g_{N-K},\bar{Z}_{1},\ldots,\bar{Z}_{K}$ form an independent
operator set. For example, suppose we find a set $\{\bar{Z}_{i}\}$
that satisfies the condition and $\{\bar{X}_{i}\}$ is the set of
corresponding logical flipping operators, it is easy to check that
$\{g_{1},\ldots,g_{N-K},\bar{X}_{1},\ldots,\bar{X}_{K}\}$ is also
an independent set. So the status of logical sets $\{\bar{Z}_{i}\}$
and $\{\bar{X}_{i}\}$ are the same, and their roles can be exchanged.

\subsection{Logical qubit encoding}

In this section, we show how to encode logical qubit in the topological
protected array. As an example, the toric code is a kind of stabilizer
code. The computation is done in the ground space of $H_{TC}=-g\sum\prod A_{s}-h\sum\prod B_{p}$,
where $A_{s}$ and $B_{p}$ are the star and plaquette operators.
The ground space is stabilized by the operators $\{A_{s},B_{p}\}$.
The loop products of $\sigma^{x}$ and $\sigma^{z}$ along two directions
give rise to the logical gate $\bar{Z}_{1}$, $\bar{Z}_{2}$, $\bar{X}_{1}$
and $\bar{X}_{2}$.

But toric code model only encodes two logical qubits. If we want to
encode more qubits, we have to implement a much too complex boundary
condition for more genus. In our system with puncture topology, we
can see that encoding is much easier.

We can write down the stabilizers in our considering model, $\{\hat{P}_{a}\}$
and $\{\hat{Q}_{abc}\}$. The physical ground space satisfying $\hat{P}_{a}|\psi\rangle=\hat{Q}_{abc}|\psi\rangle=|\psi\rangle$
constitutes the protected code space. But this is not enough. Now
we have to find the logical operators $\{\bar{Z}_{i}\}$ and $\{\bar{X}_{i}\}$.
In an array with $K$ holes, we can see that the product of $\sigma^{x}$
around the $k$th hole (i.e., $\hat{T}_{\gamma_{k}}$ mentioned earlier),
actually plays the role of logical operator $\bar{Z}_{k}$. We have
mentioned that eigenvalues $\pm1$ of $\bar{Z}_{k}$ can be regarded
as the spin values of the $k$th hole. We can also find the logical
operator $\bar{X}_{k}$ as the product of $\sigma^{z}$ along the
path from the outer boundary to the inner of the $k$th hole (The
yellow lines connecting the right hole to the outer boundary in Fig.\ref{array}).
$\bar{X}_{k}$ flips the spin of the $k$th hole to the opposite direction.

It may feel strange that we take the loop product of $\sigma^{x}$
as logical $\bar{Z}$ while taking the string product of $\sigma^{z}$
as logical $\bar{X}$. We should notice that the string product of
$\sigma^{x}$ is not allowed by constraints at the two ends of the
string. On the other hand, although we can exchange the roles of $\bar{X}$
and $\bar{Z}$ as mentioned earlier, the representation we chose here
has a better physical meaning as {}``spin''.

Therefore, we denote the logical computation basis by the spin directions
of the holes, like $|\Uparrow_{1}\Downarrow_{2}\ldots\Uparrow_{k}\ldots\rangle$.
The corresponding single qubit operations are $\bar{Z}_{k}$ and $\bar{X}_{k}$,
formally. We emphasize again that all the homotopic paths are equivalent.

\begin{figure}[ptbh]
\includegraphics[width=2.84in]{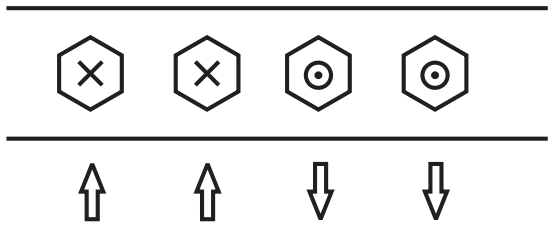} \caption{A loop product of $\sigma^{x}$ around the $k$th hole measures the
direction of its flux. The state shown is $|\Uparrow_{1}\Uparrow_{2}\Downarrow_{3}\Downarrow_{4}\rangle$. }

\label{encode}
\end{figure}

Now we show why the code space is protected against local noise. We
assume that the noise can be treated as local perturbations on the
rhombus, $\sigma_{k}^{z}$. To consider the perturbation effect of
the first order, we compute the matrix element $\langle G_{\alpha}|\sigma_{k}^{z}|G_{\beta}\rangle$.
$\{|G_{\alpha}\rangle,|G_{\beta}\rangle,\ldots\}$ is a set of orthogonal
basis of the ground space $V_{G}$, as discussed in the last section.
There must exist at least one $\hat{Q}_{abc}$ that anti-commutes
with $\sigma_{k}^{z}$, so $\sigma_{k}^{z}|G_{\beta}\rangle$ is driven
outside $V_{G}$ and $\langle G_{\alpha}|\sigma_{k}^{z}|G_{\beta}\rangle$
is zero. Only perturbations of an order higher than $N$ take effect,
such that $N$ Pauli operators form a topologically nontrivial path
across different boundaries, or form a loop embedding at least one
hole. When $N$ is large enough, the system can be well protected
and this is the reason why it is called topologically protected.

When it is cooled to a low enough temperature, the system arrives
at the protected ground space automatically. The computation is well
protected from local noises. However, proper perturbations can drive
quasi-particles tunnel through nontrivial paths, and this can be used
to construct quantum gates. We will do computations by this method
in the following part.

\subsection{Single qubit operation}

We showed how the computation basis for logical qubits are represented.
Now let us see how to implement quantum gate operations in this system.
From the discussion of the system previously, we know that quasi-particles
are excited by Pauli operators. The product operators of $\sigma^{x}$
and $\sigma^{z}$ along topologically non-trivial paths change the
state of the system. So we get a more physical picture that the transmission
of both charge and vortex changes the state. This can be used for
operation in this section.

To complete the qubit gates operation, we use a method similar to
Kou's\cite{kou09ar5}. We change the external magnetic field. Effectively,
the ground space behaves as a pseudo-spin system. We can see that
it just drives the quasi-particles to tunnel through different sectors.

As an example, we slightly tune the flux of the rhombi along a path
from the $k$th hole to the outer boundary. According to the previous
discussion, perturbations of the $\sigma^{z}$ form appear in the
single-rhombus Hamiltonian, and this is allowed by the constraint.
So we get the Hamiltonian \begin{equation}
\hat{\mathcal{H}}=\hat{\mathcal{H}}_{0}+\frac{\Delta E}{2}\sum_{path}\sigma^{z}=\hat{\mathcal{H}}_{0}+H'\end{equation}

Confined in the degenerate ground space. the main part gives identity
with an energy coefficient. The effect of perturbation appears in
the $(N-1)$th order. $N$ is the number of rhombi along the path.
One can get the energy splitting\cite{kou09ar5} \begin{equation}
\delta\epsilon=\sum_{j}\langle\Uparrow_{k}|H'(\frac{1}{E_{0}-\hat{\mathcal{H}}_{0}}H')^{j}|\Downarrow_{k}\rangle=\frac{(\Delta E/2)^{N}}{(-4r)^{N-1}},\label{split}\end{equation}
 where $E_{0}$ is the ground energy of $\hat{\mathcal{H}}_{0}$.
Each $H'$ contributes one single $\sigma^{z}$. Only the term that
contains a complete string product of $\sigma^{z}$ connecting both
the inner and outer boundaries takes effect as $\bar{X}_{k}$ (remember
that in the representation we use here, products of $\sigma^{z}$
behave as $\bar{X}$ not $\bar{Z}$!), while others vanish. So we
get the effective Hamiltonian of the array as $\hat{\mathcal{H}}_{\textbf{eff}}^{x}=\delta\epsilon\bar{X}_{k}$.

In this process, a charge tunnels across the array from the $k$th
hole to the outer boundary. Degenerated states split just as in the
case in the double-well of a single rhombus.

Similarly, we want to get an effective Hamiltonian like $\hat{\mathcal{H}}_{\textbf{eff}}^{z}\sim\bar{Z}_{k}$
by adding perturbations of the form such as $\sum\sigma^{x}$. However,
as mentioned before, this is not allowed by the constraint under ideal
condition. Something more should be done. We have talked about the
vortex excitation in the last section. To achieve this purpose, we
carefully change the flux of the hexagons along the loop around the
$k$th hole (as shown in Fig.\ref{array}) to be integer multiples
of $\Phi_{0}$. The constraint is broken, and the dynamics of the
corresponding hexagon changes. The single flip of a rhombus is allowed.
So we get the desired perturbation as \begin{equation}
\hat{\mathcal{H}}=\hat{\mathcal{H}}_{0}+\tilde{t}\sum_{loop}\sigma^{x},\end{equation}
 where $\tilde{t}$ takes the value in Eq.(\ref{rhoH}). Two vortices
emerge. One is transmitted around the $k$th hole, and then annihilates
the other one. This process gives the $\bar{Z}$ term. Similar to
Eq.(\ref{split}), we get the energy splitting \begin{equation}
\delta E=\sum_{j}\langle\Uparrow_{k}|H'(\frac{1}{E_{0}-\hat{\mathcal{H}}_{0}}H')^{j}|\Uparrow_{k}\rangle=\frac{\tilde{t}^{N}}{(-2E_{g})^{N-1}}.\end{equation}
 So we have got $\hat{\mathcal{H}}_{\textbf{eff}}^{z}=\delta E\bar{Z}_{k}$.

As we known, any single-qubit rotation can be written as \begin{equation}
U(\theta,\phi,\gamma)=e^{-i\gamma\bar{Z}}e^{-i\phi\bar{X}}e^{-i\theta\bar{Z}}.\end{equation}
 We have got both $\hat{\mathcal{H}}_{\textbf{eff}}^{z}=\delta\epsilon\bar{Z}_{k}$
and $\hat{\mathcal{H}}_{\textbf{eff}}^{x}=\delta E\bar{X}_{k}$, then
we can get any single qubit gate, at least in principle%
\footnote{In practice, the evolution time is controlled by a electric pulse
with specific length, but usually we cannot produce pulses with \emph{arbitrarily
continous} length. But we can still complete arbitrary single qubit
operation approximately with a limited set of gates, so we just need
several specific pulses.%
}.

We can tune the corresponding flux of rhombi or hexagons and get the
effective Hamiltonian we need, carefully control the evolution time
of the system (that is to tune the parameters $\gamma,\phi,\theta$
above), then we can construct any single qubit gate we want.

\begin{figure}[!hptb]
\includegraphics[width=2.84in]{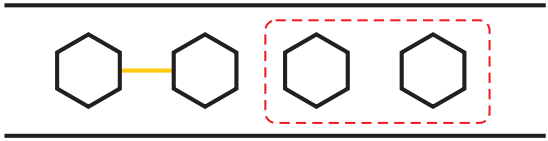} \caption{(Color online) Multi-qubit operation is sketched here. The (yellow)
solid line represents the rhombi with flux tuned by $\delta\Phi$.
This gives the effective Hamiltonian $\hat{\mathcal{H}}_{\textbf{eff}}^{xx}=\alpha\bar{X}_{m}\otimes\bar{X}_{n}$.
The red dashed loop represents the hexagons with flux changed to integer
multiples of $\Phi_{0}$. That gives $\hat{\mathcal{H}}_{\textbf{eff}}^{zz}=\beta\bar{Z}_{m}\otimes\bar{Z}_{n}$. }

\label{multi}
\end{figure}

\subsection{Multi-qubit operation and CNOT gate}

In the last part, we saw that any single qubit gate can be constructed.
In the this section, we show that in this array multi-qubit operations
are also easy to realize on any two qubits.

When we change the flux of the rhombi on the line connecting two holes
(see the yellow line in Fig.\ref{multi}), similarly, the states of
the corresponding two qubits are flipped at the same time, and this
gives $\bar{X}_{m}\otimes\bar{X}_{n}$ term in the Hamiltonian. By
using the perturbation approach, the amplitude is \begin{equation}
\alpha=\sum_{j}\langle\Uparrow_{m}\Uparrow_{n}|H'(\frac{1}{E_{0}-\hat{\mathcal{H}}_{0}}H')^{j}|\Downarrow_{m}\Downarrow_{n}\rangle=\frac{(\Delta E/2)^{N}}{(-4r)^{N-1}}.\end{equation}
 Here, $H'=\frac{\Delta E}{2}\sum\sigma^{z}$, and the summation contains
all the rhombi on yellow line. We get a effective Hamiltonian $\hat{\mathcal{H}}_{\textbf{eff}}^{xx}=\alpha\bar{X}_{m}\otimes\bar{X}_{n}$.

Analogously, we can get another type of operation. We can tune the
flux of the hexagons along the loop around two holes to integer multiples
of $\Phi_{0}$. This gives a perturbation $H'=\tilde{t}\sum\sigma^{x}$.
The amplitude is \begin{equation}
\beta=\sum_{j}\langle\Uparrow_{m}\Uparrow_{n}|H'(\frac{1}{E_{0}-\hat{\mathcal{H}}_{0}}H')^{j}|\Uparrow_{m}\Uparrow_{n}\rangle=\frac{\tilde{t}^{N}}{(-2E_{g})^{N-1}}.\end{equation}
 Then we get the effective Hamiltonian $\hat{\mathcal{H}}_{\textbf{eff}}^{zz}=\beta\bar{Z}_{m}\otimes\bar{Z}_{n}$.

The most important thing we are concerned with is to construct CNOT
operation, because CNOT gate together with arbitrary single qubit
rotation gate gives rise to a universal quantum computation. This
is also realizable in this array.

First, we should construct another effective Hamiltonian. We tune
the flux in such a way that a vortex is transmitted around the $m$th
hole, and a charge is transmitted from the $n$th hole to the outer
boundary at the same time. The effective Hamiltonian contains contributions
of three parts, $\hat{\mathcal{H}}_{\textbf{eff}}^{zx}=\gamma_{1}\bar{Z}_{m}+\gamma_{2}\bar{X}_{n}+\gamma_{3}\bar{Z}_{m}\otimes\bar{X}_{n}$.
$\gamma_{1},\gamma_{2}$ and $\gamma_{3}=\gamma_{1}\gamma_{2}$ are
evaluated in the same way with $\delta E$ and $\delta\epsilon$ in
the last section.

A CNOT gate can be constructed with the help of $\hat{\mathcal{H}}_{\textbf{eff}}^{zx}$,
$\hat{\mathcal{H}}_{\textbf{eff}}^{z}$ and $\hat{\mathcal{H}}_{\textbf{eff}}^{x}$.
We can let $\hat{\mathcal{H}}_{\textbf{eff}}^{zx}$ evolve for $t_{1}=\frac{\pi}{4\gamma_{3}}$.
Then we should change the external magnetic field to construct $\hat{\mathcal{H}}_{\textbf{eff}}^{z}$
and let the system evolve for $t_{2}=\frac{3\pi}{4\gamma_{1}}-t_{1}$.
Finally, we construct $\hat{\mathcal{H}}_{\textbf{eff}}^{x}$ and
let the system evolve for $t_{3}=\frac{\pi}{2\gamma_{2}}-t_{1}$.
Now we get the CNOT operation with an external global phase $e^{-i\pi/4}$,
which is negligible. \begin{equation}
e^{-i\pi/4}U_{\text{CNOT}}=e^{-i\hat{\mathcal{H}}^{zx}t_{1}}e^{-i\hat{\mathcal{H}}^{z}t_{2}}e^{-i\hat{\mathcal{H}}^{x}t_{3}}\end{equation}

\section{Summary}

In summary, we propose a TQC scheme with protected qubits in a JJA
system, and we indicate how to perform qubit encoding, single qubit
operation, and multi-qubit operation. Especially, we show the way
to implement the CNOT quantum gate operation which is a key element
in quantum computation.

The scheme we propose here can be realized in real JJA system since
the technology of the Josephson junction is developing rapidly. Considering
the convenience of realizability and scalability of our system, the
dimension of ground space is related to the number of punctures on
the array. The logic qubit encoding here is in a protected space so
that our system is topologically protected from noise. Besides, specific
quasiparticles are driven by proper perturbation transmitted along
topologically nontrivial path. This makes small splittings happen.
We give the effective Hamiltonian under different condition corresponding
to various splitting situations and this can be used to do quantum
computation.

The work is supported in part by the NSF of China Grant Nos. 90503009
and 10775116 and 973-Program Grant No. 2005CB724508.

\end{document}